\documentstyle[aps,preprint]{revtex}
\input epsf.sty

\begin{document}
\draft
\flushbottom

\title{Tunneling spectroscopy in the magnetic superconductor TmNi$_2$B$_2$C}
\author{H. Suderow${^1}$, P. Martinez-Samper${^2}$, S. Vieira${^2}$
\\
N. Luchier${^3}$, J.P. Brison${^3}$
\\
P. Canfield${^4}$}
\address{
${^1}$ Instituto de Ciencia de Materiales de Madrid (ICMM) \\
Consejo Superior de Investigaciones Cientificas \\ Campus de
Cantoblanco 28049 Madrid-Spain\\
${^2}$ Laboratorio de Bajas Temperaturas, Departamento de Fisica de
la Materia Condensada \\ Instituto de Ciencia de Materiales
Nicol\'as Cabrera, Facultad de Ciencias, C-III \\ Universidad
Aut\'onoma de Madrid, 28049 Madrid-Spain
\\
  ${^3}$ Centre des Recherches sur
les Tres Basses Temperatures \\ CNRS, BP 166, 38042 Grenoble Cedex
9, France
\\
${^4}$ Ames Laboratory and Departament of Physics and Astronomy \\
Iowa State University, Ames, Iowa 50011 }
\date{\today}
\maketitle


\begin{abstract}
We present new measurements about the tunneling conductance in the
borocarbide superconductor TmNi$_2$B$_2$C. The results show a very
good agreement with weak coupling BCS theory, without any lifetime
broadening parameter, over the whole sample surface. We detect no
particular change of the tunneling spectroscopy below 1.5K, when both
the antiferromagnetic (AF) phase and the superconducting order coexist.
\end{abstract}



Electron tunneling spectroscopy is a very powerful tool to
investigate the fundamental properties of conducting materials
which has been widely applied to the study of superconductivity
\cite{Wolf}. The results have verified several of the most
important predictions of the BCS theory. More recently, Scanning
Tunneling Microscopy and Spectroscopy (STM/STS) has given the
possibility to measure local variations of the superconducting
density of states. For example, the vortex structure of several
superconductors, such as NbSe$_2$, some of the High T$_c$ oxydes or
the borocarbide superconductors, could be studied in
detail\cite{Hess90,Maggio95,Pan00,Sakata00,deWilde97}.
Nevertheless, it remains very difficult to get accurate information
from the most simple measurement one can do with STM/STS, the zero
field tunneling conductance curves. Indeed, the fitting of the
measured curves by any kind of model requires the introduction of
an ad'hoc broadning parameter \cite{Dynes78} which is supposed to
account phenomenologically for uncharacterized pair-breaking
effects, and whose magnitude is often comparable to that of the gap
itself. So only the zero field curves obtained in the most simple
superconductors (as e.g. Pb, Nb or Al
\cite{Pan00b,Suderow00,Yazdani97,NbSe2}) have been successfully
fitted by BCS or Eliashberg theory. The broadening parameter
appears also when data are be obtained using other related
techniques such as break-junctions, point contact spectroscopy, or
planar junctions in the more complex materials (as for example
heavy fermions or
borocarbides\cite{Jourdan99,Loehneysen96,Ekino96,Yanson00}).

Here we present new data in TmNi$_2$B$_2$C. This compound belongs
to the family of the borocarbide superconductors (RENi$_2$B$_2$C
where RE is a rare earth; RE = Lu, Y, Tm, Er,
Ho\ldots)\cite{Canfieldgeneral,CavaNagarajan,Lynn97,Cho95}. These
systems present rich phase diagrams showing coexistence or
competition between superconductivity and a magnetic order of the
RE spins (when they have a magnetic moment)\cite{Cho95}. In the Tm
compound the superconducting and Neel critical temperatures are
well separated ($T_c=10.5K$, and
$T_N=1.5K$)\cite{Movshovich94a,Lynn97,Cho95} and in the
antiferromagnetic phase, the spins of the Tm ions order in a
transversely polarized spin density wave with an incommensurate
modulation of the magnetic moments \cite{Lynn97,Norgaard00}. To our
knowledge, no previously published tunneling experiment is
available in this compound. The tunneling experiments reported up
to now in nonmagnetic compounds and the point contact spectroscopy
data in magnetic compounds
\cite{deWilde97,Sakata00,Ekino96,Yanson00} have never shown a
simple BCS density of states $N_{BCS}(E)$ and the conductance
curves were all severely broadened. Therefore, the possible
presence of low energy excitations as well as the form of the
superconducting gap in the coexistence region were masked by
extrinsic broadening. Here we present tunneling spectra that do not
show any additional broadening, even in the antiferromagnetic
phase. These data present a significant advance in tunneling
spectroscopy and they should allow further progress on the detailed
study of the superconducting ground state of borocarbides.

We have used a STM with an x-y table that permits coarse mouvement
in a 2x2 mm$^{2}$ region in a $^{3}$He insert, and where the sample
holder is cooled down to 0.8K. The tip is prepared from a gold wire
cut with a clean blade. We measured several samples (of dimensions
about 1x1x1 mm$^{3}$), prepared by breaking the same single crystal
platlet grown by a flux technic described in
\cite{Canfieldgeneral,Canfield}. The best results were obtained
with the samples which were mounted on the STM and cooled down
immediately after breaking, so that the surface remained no more
than several minutes at ambient pressure (about 10 in most cases).
They show good quality, highly reproducible spectra over the whole
surface.
  The topographic images (Fig.1) are always of good quality, and
independent of the tunneling resistance. They show an irregular
structure with cluster-like forms, with a diameter of the order of
20-30 nm on planes of about 30-60 degrees. In spite of intensive
search with the x-y table, we were unable to find sufficiently flat
zones to obtain atomic resolution. These results are similar to the
observations made in other borocarbides\cite{Sakata00}.

It is well known that proper RF filtering is essential to do
tunneling or point contact spectroscopy in
superconductors\cite{Wolf}. This is specially important in an STM
set-up as the tunneling current is of the order of nA or lower (we
use tunneling resistances between $1M\Omega$ and $10M\Omega$): this
is several orders of magnitude smaller than in a normal planar
junction experiment. In TmNi$_2$B$_2$C, we have found an important
difference between measurements obtained at the same temperature
and with the same surface preparation but in two different
cryostats. The same low noise STM electronics was used on both
setup but one had filters and the other none. In the unfiltered
setup we find a fictitious finite conductance at zero bias and we
need to introduce a broadening parameter $\Gamma\approx
0.3\Delta$\cite{Dynes78}, comparable to the values obtained in
previous works \cite{deWilde97,Sakata00,Ekino96,Yanson00}. On the
contrary, the curve measured with appropriate shielding can be
accurately fitted to the conventional BCS expression, using the
superconducting gap $\Delta$ and the temperature $T$ as the only
fitting parameters. We use room temperature RF feedthrough filters
and thermocoaxial cables for all the electrical connections of the
STM. The results presented in this paper have been verified on
three different samples measured in the filtered set-up, two of
them with the surface parallel to the a-b plane, and one of them
with the surface parallel to the c axis of the tetragonal crystal
structure. No significant differences were found. Note nevertheless
that the surface (Fig.1) consists of inclined planes that do not
correspond to a clear crystallographic direction.

For the curve in Fig.2a, the best fit to the experiment (line on
Fig.2a ) is obtained with $\Delta=1.4mV$ and $T=2K$.  The upper
limit to any broadening parameter is ($\Gamma < 0.001\Delta$). Note
that the temperature obtained from the fit is larger, by 0.2 K,
than the actual sample temperature.  Indeed, the quasiparticle
anomaly at $\Delta$ is not as peaked as expected from BCS theory
and adding a finite $\Gamma$ does not give an appropriate fit, as
it immediately results in increasing the zero bias conductance.  It
could be that we still need a better filtering, but our
measurements in Pb and Al samples down to lower temperatures (400
mK \cite{nosotrosBT}) have shown that our (relative) spectral
resolution is sufficiently good to resolve such details.  A more
likely explanation is that the superconducting gap has a small
anisotropy.  Indeed, a marked anisotropy or the presence of
different gaps results in the appearance of different maxima in the
density of states at $T=0K$ and therefore in a broadened
quasiparticle peak at finite temperature, without adding additional
conductance at zero bias\cite{Wolf,NbSe2}.  For instance, the
measured tunneling conductance in the superconductor NbSe$_2$ shows
a quasiparticle anomaly that is much less peaked than expected from
BCS theory.  It also has a marked structure possibly corresponding
to the very large anisotropy of the superconducting gap found in
that compound \cite{Hess90,Pan00,NbSe2}.  In our case, the decrease
in peak intensity is much smaller so that a slight dispersion in
the value of the measured gap of the order of several percent of
$\Delta$ could already account for the observed broadening.
However, a quantitative calculation is by now not realistic,
because it requires knowledge of the different values of the gap
$\Delta(k)$ for the relevant set of $k$ vectors.

Let us note that there are several other experimental indications
of gap anisotropies, or equivalently of multiband (each having a
proper gap amplitude) effects in borocarbide superconductors.  The
upward curvatures of the upper critical field in LuNi$_{2}$B$_{2}$C
and YNi$_{2}$B$_{2}$C also present in TmNi$_{2}$B$_{2}$C (for H in
the basal plane) have been explained first by non-local effects
\cite{Metlushko97}, backed by the structural changes in the vortex
lattice \cite{McPaul 98,Eskildsen97,Eskildsen97b,Cheon98,Gammel99}.
But an alternate (or complementary) interpretation to the positive
curvatures involves a two band model with different coupling
constants and so different gap amplitudes \cite{Shulga98}.  The
T$^{3}$ dependence of the zero field electronic part of the
specific heat at low temperatures in YNi$_{2}$B$_{2}$C or
LuNi$_{2}$B$_{2}$C, together with its strong field dependence could
also point toward reduced gap regions on the Fermi surface
\cite{Nohara99}, although our measurements indicate that the
reduction of the gap in these regions and in the case of Tm is not
larger than several percent.

When cooling down towards the antiferromagnetic phase ($T_N=1.5K$),
we do not observe any change in the spectroscopy, other than the
temperature reduction (in Fig.2b we use $T=1K$ and
$\Delta=1.45mV$).  What is more, the whole temperature dependence
of the superconducting gap shown in Fig.3 is in good agreement with
conventional BCS theory, within experimental error. The value of
the superconducting gap gives $\Delta/k_BT_c=1.55$ (taking
$\Delta=1.45mV$ and $T_c=10.5K$).  Due to our error bars on the
determination of $T_c$ ($\pm 15 \%$) this ratio is in agreement
with the BCS weak coupling value of $1.73$.  This value raises the
same problems as those discussed in Ref. \cite{Ekino96}. In this
work, tunneling spectroscopy has been performed on the Y and Lu
compounds with break-junctions.  The curves show significant
broadening, and yield a large range of ratios of $\Delta/k_BT_c$
(depending on the surface or junction quality).  But the
weak-coupling regime was put forward because these ratios are
systematically lower or equal than the BCS value. Our results on
the Tm compound also strongly support weak coupling
superconductivity, thanks both to the absence of broadening and to
the reproducibility of the results over very large surface areas.

On the other hand, analysis of the specific heat jump at T$_{c}$,
and of the low temperature regime of the specific heat, implies a
rather small ratio of $T_{c}/\omega$ where $\omega$ is some average
frequency of the phonon spectrum\cite{Wolf}, meaning strong to
intermediate coupling.  Typically, for the Y and Lu compound,
values for $\omega$ between $150K$ and $200K$ are needed to explain
the specific heat results \cite{Michor95}. They imply values of
$\Delta/k_BT_c\approx 2$ that are clearly incompatible to the value
we have found in Tm, which is not expected to have a very different
phonon spectrum. In addition, $\omega$ between $150K$ and $200K$
implies the appearance of anomalies characteristic for strong
coupling superconductivity clearly within our experimental
resolution on the tunneling density of states between $13$ and
$17mV$\cite{Wolf}, which neither we nor the authors of
Ref.\cite{Ekino96} have observed. By contrast $\omega\approx 490K$,
proposed for the two band model of the upper critical field in Ref.
\cite{Shulga98}, fits much better our results. But this leaves the
problem of the discrepancy with the thermodynamic data unsolved.
The solution to these contradictions may lie in the complex phonon
spectrum (see references and a model spectrum in \cite{Michor95}),
and again on the multiband structure of the Fermi surface in
borocarbide superconductors. They may yield different results for
the average frequency involved in the specific heat, the gap, or
the upper critical field\cite{Carbotte90}. This looks somewhat
paradoxical in these compounds which, except for their magnetic
properties, show rather isotropic behavior in the normal phase.

As regard the lack of signature of the antiferromagnetic order in
the tunneling spectra, it might be explained more intuitively.  For
example, one could invoke that the period of the incommensurate
magnetic modulation is of the order of $2.5$nm, which is small
compared to the superconducting coherence length $\xi_{0} = 12$
nm\cite{Cho95,Eskildsen98}.  This means that the variation of the
local magnetic moment is averaged out on the superconducting
coherence length or, in q-space, that the superconducting order
does not make any change on the susceptibility $\chi(q)$ at the
antiferromagnetic wave-vector $Q>>1/\xi_{0}$.  In order to test
more exotic possibilities, like a local depression of the
superconducting gap or the presence of new magnetic excitations of
longer period, not detected by other techniques, we made local
spectroscopy measurements. An I-V curve is done at a given set of
pixels in a topography image.
 From the difference $dI/dV(V>\Delta)-dI/dV(V<\Delta)$ we can test the
local appearence of low energy excitations.  But we obtained
essentially flat images giving $dI/dV(V>\Delta)-dI/dV(V<\Delta)=1$
within $5\%$, as shown in Fig.4 for a $240$x$240nm$ scan.

We do not know of any previous measurements of the tunneling
conductance in an antiferromagnetic superconductor. Indeed, it is
well known, mainly through macroscopic measurements as resistivity
or specific heat, that an antiferromagnetic order coexists with
superconductivity in this compound \cite{Cho95,Movshovich94a}. But
their mutual influence is a hotly debated issue (see
\cite{Norgaard00} and \cite{Bulaevski85}) and the predictions
remained ambiguous. For instance, the authors of Ref.\cite{Kulic97}
point towards the possible existence of a line of nodes in the
superconducting gap in the antiferromagnetic phase of
TmNi$_2$B$_2$C. We have now ruled out this possibility.

In conclusion, we have studied the superconducting gap of
TmNi$_2$B$_2$C as a function of temperature. The results can be
fitted to conventional, BCS, weak-coupling theory in the whole
temperature range. We did not detect any change of the
superconducting phase when the antiferromagnetic order appears.

\begin{figure}
\epsfxsize 12cm
\epsfbox{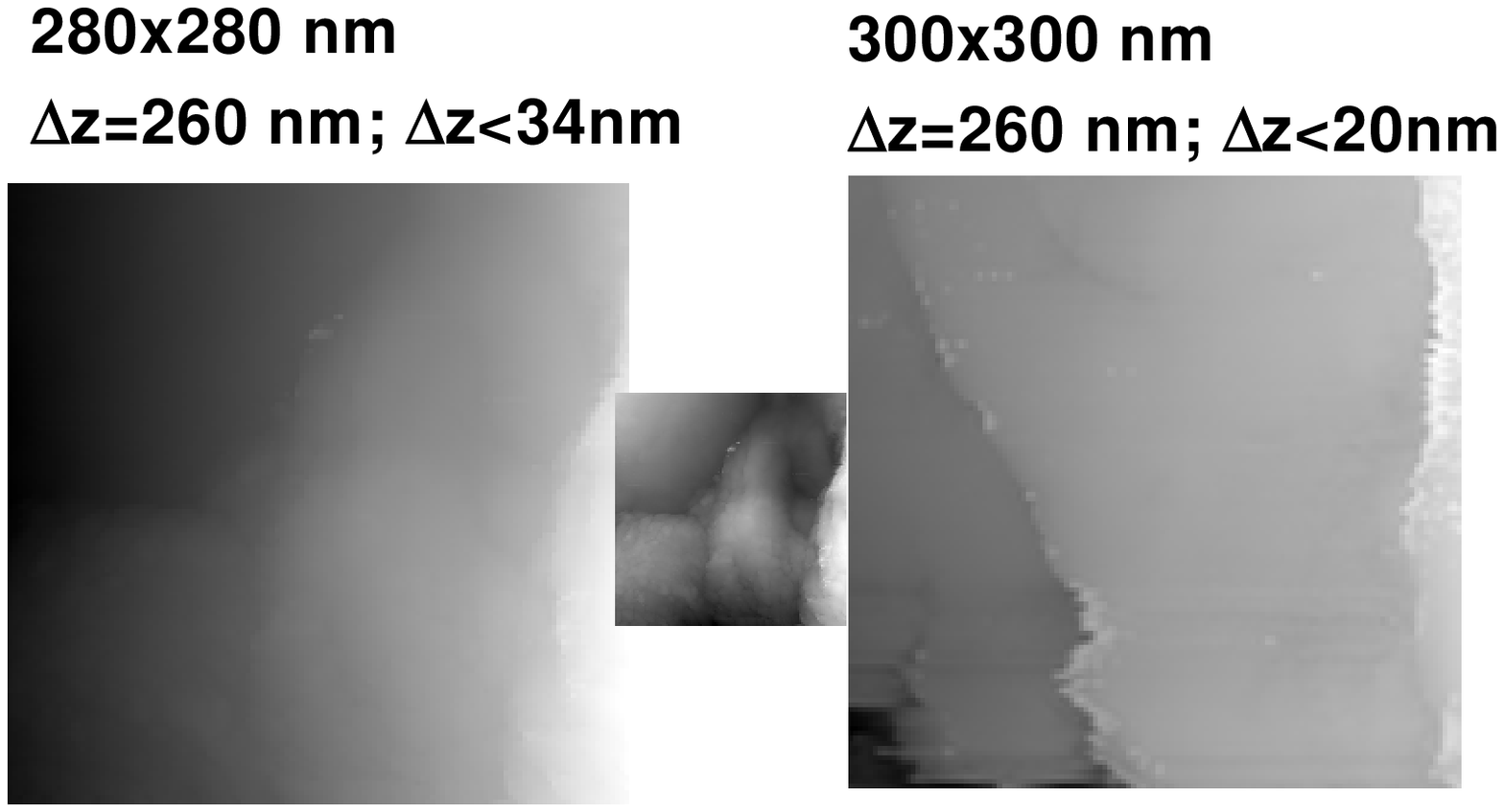}
\caption{
The figure shows typical topographic images on TmNi$_2$B$_2$C. Sometimes
terraces of some tens nm height can be observed. The typical image
is however an inclined surface with "bumps" of 20-30 nm in height.
The bumps are visualized in the figure between both images.}
\label{fig:Fig1}
\end{figure}

\begin{figure}
\epsfxsize 12cm
\epsfbox{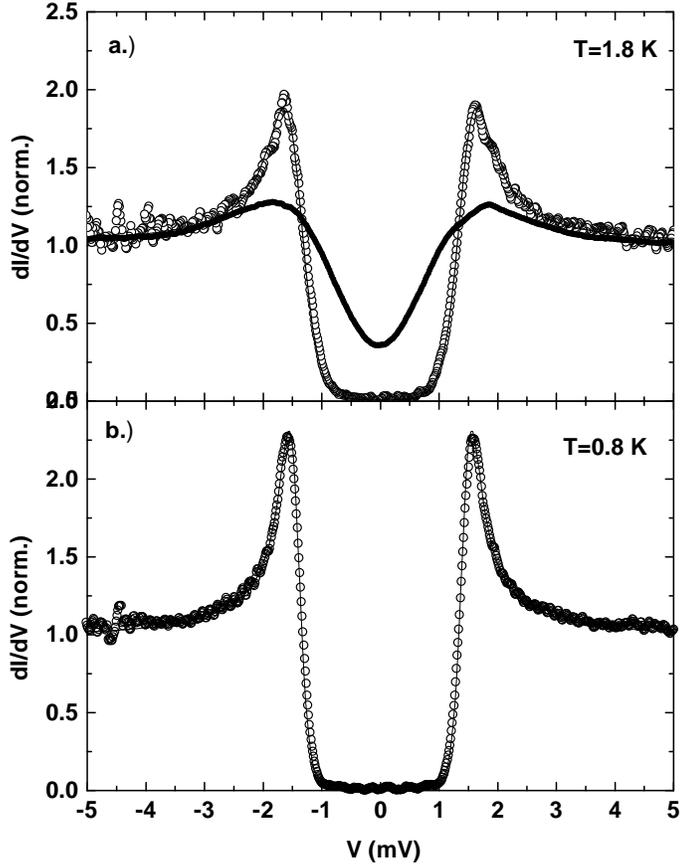}
\caption{
The superconducting gap at 1.8 and 0.8 K as measured with tunneling
spectroscopy (tunneling resistance about $10M\Omega$). In a. we
show the important improvement of the quality of the I-V curves
between the measurements done in an unfiltered (full points) and in
a filtered setup (open circles). In b. we show the result in the
Antiferromagnetic phase (T$_N$=1.5K). No changes are observed. The
lines are fits to the BCS weak coupling theory using the parameters
given in the figures.}
\label{fig:Fig2}
\end{figure}

\begin{figure}
\epsfxsize 12cm
\epsfbox{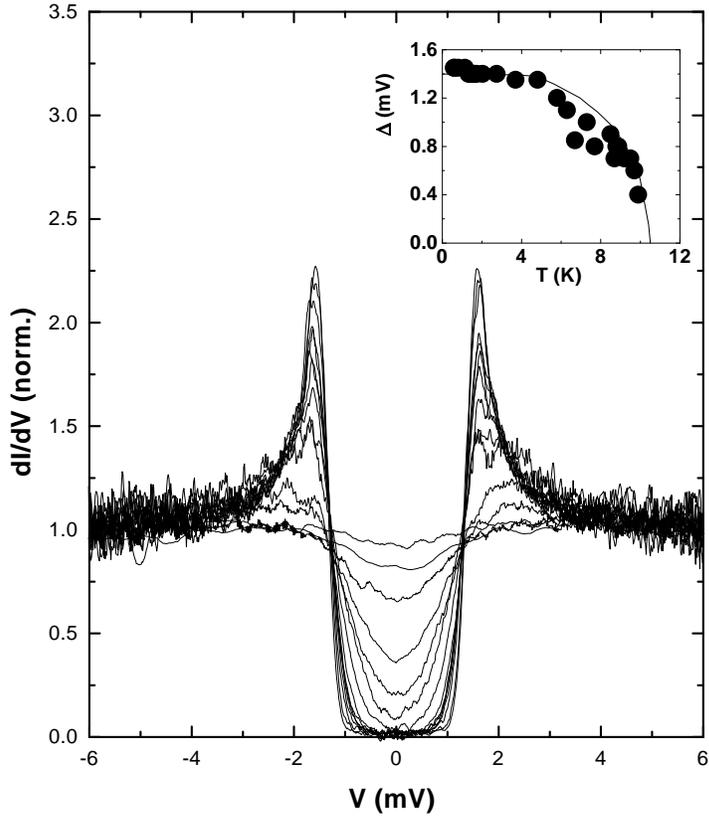}
\caption{
The temperature dependence of the superconducting gap as followed
by tunneling spectroscopy. Fitting the experimental curves gives
the gap shown in the inset. The scattering in the results increases
above 6K, possibly due to the experimental uncertainity.}
\label{fig:Fig3}
\end{figure}

\begin{figure}
\epsfxsize 12cm
\epsfbox{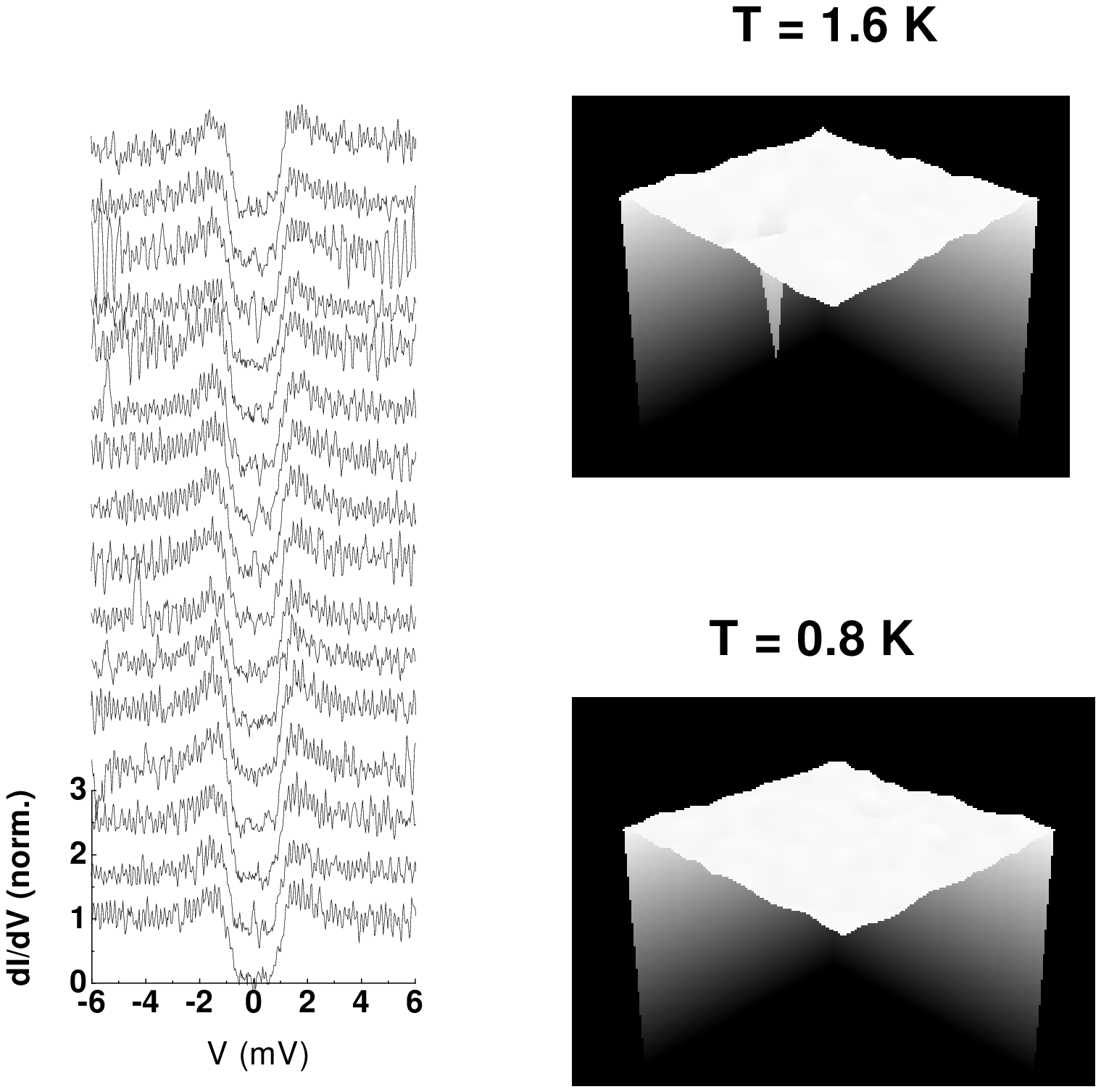}
\caption{
Here we show two images done by doing an I-V sweep at a set of
16x16 points in a surface of 240x240 nm. Note that no changes are
observed below and above the Neel temperature. Some of the dI/dV(V)
curves are shown in the left side. Note that the experimental
resolution is less as compared to the curves shown in  Figs.2 and
4. order to obtain the image in a reasonable time (45 minutes).}
\label{fig:Fig4}
\end{figure}


\begin{references}

\bibitem{Wolf}
E.L. Wolf, "Principles of Electron Tunneling Spectroscopy", Oxford
University Press (1989).

\bibitem{Hess90}
H.F. Hess, R.B. Robinson, J.V. Waszcak, Phys. Rev. Lett. {\bf 64},
p. 2711 (1990).

\bibitem{Maggio95}
I. Maggio-Aprili, Ch. Renner, A. Erb, E. Walker, O. Fisher, Phys.
Rev. Lett. {\bf 64}, p. 2711 (1990).

\bibitem{Pan00}
S.H. Pan, E.W. Hudson, K.M. Lang, H. Eisaki, S. Uchida, J.C. Davis,
Nature {\bf 403}, p. 746 (2000).

\bibitem{Sakata00}
H. Sakata, M. Oosawa, K. Matsuba, N. Nishida, H. Takeya, K. Hirata,
Phys. Rev. Lett. {\bf 84}, p. 1583 (2000).

\bibitem{deWilde97}
Y. De Wilde, M. Iavarone, U. Welp, V. Metlushko, A.E. Koshelev, I.
Aranson, G.W. Crabtree, P.C. Canfield, Phys. Rev. Lett. {\bf 78},
p. 4273 (1997).

\bibitem{Dynes78}
R.C. Dynes, V. Narayanamurti, J.P. Garno, Phys. Rev. Lett., {\bf
41},p. 1509 (1978).

\bibitem{Pan00b}
S.H. Pan, E.W. Hudson, J.C. Davis, Applied Phys. Lett., {\bf 73},
p. 2992 (1998).

\bibitem{Suderow00}
H. Suderow, A. Izquierdo, S. Vieira, Physica C, {\bf 332}, p. 327
(2000).

\bibitem{Yazdani97}
A. Yazdani, B.A. Jones, C.P. Lutz, M.F. Crommie, D.M. Eigler,
Science, {\bf 275}, 1767 (1997).

\bibitem{NbSe2}
P. Martinez Samper, E. Bascones, H. Suderow, J.G. Rodrigo, F.
Guinea and S. Vieira, to be published.

\bibitem{Jourdan99}
M. Jourdan, M. Huth, H. Adrian, Nature {\bf 398}, 47 (1999).

\bibitem{Loehneysen96}
See e.g. H.v. Loehneysen, Physica B, {\bf 218}, p.148 (1996) and
References therein.

\bibitem{Ekino96}
T. Ekino, H. Fujii, M. Kosugi, Y. Zenitani, J. Akimitsu, Phys. Rev.
B, {\bf 53}, p. 5640 (1996).

\bibitem{Yanson00}
I.K. Yanson, N.L. Bobrov, C.V. Tomy, D. McK. Paul, Physica C, {\bf
334}, p.33 (2000).

\bibitem{Canfieldgeneral}
P.C. Canfield, P.L. Gammel, D.J. Bishop, Phys. Today (1998), {\bf
51}, pp. 40-46.

\bibitem{CavaNagarajan}
R.J. Cava et al. Nature, {\bf 367}, 146 (1994); R. Nagarajan et al.
Phys. Rev. Lett. {\bf 72}, 274 (1994).

\bibitem{Lynn97}
J.W. Lynn, S. Skanthakumar, Q. Huang, S.K. Sinha, Z. Hossain, L.C.
Gupta, R. Nagarajan, C. Godart, Phys. Rev. B, {\bf 55}, p. 6584
(1997).

\bibitem{Cho95}
B.K. Cho, M. Xu, P.C. Canfield, L.L. Miller, D.C. Johnston, Phys.
Rev. B, {\bf 52}, p. 3676 (1995)


\bibitem{Movshovich94a}
R. Movshovich, M.F. Hundley, J.D. Thompson, P.C. Canfield, B.K. Cho
and A.V. Chubukov, Physica C, {\bf 227}, 381-6 (1994)

\bibitem{Norgaard00}
K. Norgaard, M.R. Eskildsen, N.H. Andersen, J. Jensen, P. Hedegard,
S.N. Klausen, P.C. Canfield, Phys. Rev. Lett. {\bf 84}, p. 4982
(2000).

\bibitem{Canfield}
B. K. Cho, P. C. Canfield, L. L. Miller, D. C. Johnston, W. P.
Beyermann and A. Yatskar
, Phys. Rev. B {\bf 52}, p. 3684 (1995)

\bibitem{nosotrosBT}
H. Suderow, S. Vieira, to be published.

\bibitem{Metlushko97} V. Metlushko, U. Welp, A. Koshelev, I. Aronson,
G.W.Crabtree and P.C. Canfield, Phys.  rev.  Lett.  {\bf 79}, 1738
(1997).

\bibitem{McPaul 98} D. McK. Paul, C.V. Tomy, C.M. Aegerter, R. Cubitt,
S.H. Lloyd, E.M. Forgan, S.L. Lee and M. Tethiraj, Phys.  Rev.  Lett.
{\bf 79}, 1738 (1997).

\bibitem{Eskildsen97}
M.R. Eskildsen, P.L. Gammel, N.P. Barber, U. Yaron, A.P. Ramirez,
D.A. Huse, D.J. Bishop, C. Bolle, C.M. Lieber, S. Oxx, S. Sridhar,
 N.H. Andersen, K. Mortensen, P.C. Canfield Phys. Rev. Lett. (1997),
{\bf 78},  1968-1971.

\bibitem{Eskildsen97b}
M.R. Eskildsen, P.L. Gammel, B.P. Barber, A.P. Ramirez, D.J.
Bishop, N.H. Andersen, K. Mortensen, C.A. Bolle, C.M. Lieber, P.C.
Canfield, Phys. Rev. Lett.  (1997),  {\bf 79},  487-490.

\bibitem{Cheon98}
K.O. Cheon, I.R. Fisher, V. Kogan, P.C. Canfield, P. Miranovic,
P.L. Gammel, Phys. Rev. B: Condens. Matter Mater. Phys.  (1998),
{\bf 58}, 6463-6467.

\bibitem{Gammel99}
P.L. Gammel, D.J. Bishop, M.R. Eskildsen, K. Mortensen, N.H.
Andersen, I.R. Fisher, K.O. Cheon, P.C. Canfield, V.G. Kogan, Phys.
Rev. Lett.  (1999),  {\bf 82},  4082-4085

\bibitem{Shulga98} S.V. Shulga, S.-L. Drechsler, G. Fuchs, K.H.
M\"{u}ller, K. Winzer, M. Heinecke and K. Krug, Phys.  Rev.  Lett.
{\bf 80}, 1730 (1998).

\bibitem{Nohara99} M. Nohara, M. Isshiki, F. Sakai and H. Takagi, J.
Phys. Soc. Jpn. {\bf 68}, 1078 (1999).

\bibitem{Michor95} H. Michor, T. Holubar, C. Dusek and G. Hilscher,
Phys. Rev. B {\bf 52}, 16165 (1995).

\bibitem{Carbotte90}
Carbotte J.P., Rev. Mod. Phys. {\bf 62}, 1027 (1990).

\bibitem{Eskildsen98}
M.R. Eskildsen, K. Harada, P.L. Gammel, A.B. Abrahamsen, N.H.
Andersen, G. Ernst, A.P. Ramirez, D.J. Bishop, K. Mortensen, D.G.
Naugle, K.D.D. Rathnayaka, P.C. Canfield, Nature {\bf 393}, p. 242
(1998).

\bibitem{Bulaevski85}
L.N. Bulaevski, A.I. Buzdin, M.L. Kulic, S.V. Panjukov, Adv. in
Phys. (1985), {\bf 34}, p. 175.

\bibitem{Kulic97}
M.L. Kulic, A.I. Buzdin, L.N. Bulaevskii, Phys. Lett. A, p. 285
{\bf 235} (1997)



\end{references}
\end{document}